\documentclass[10pt,english,manuscript]{revtex4}
\usepackage[T1]{fontenc}
\usepackage[latin9]{inputenc}
\usepackage{amsmath}
\usepackage{amssymb}
\usepackage{graphicx}

\makeatother
\usepackage{babel}

\begin{document}

\title{Magnetic properties, Lyapunov exponent and superstability of
the spin-$\frac{1}{2}$ Ising-Heisenberg model on diamond chain}

\author{N.S. Ananikian$^{1,2}$ and V.V. Hovhannisyan$^{1}$}

\affiliation{$^{1}$A.I. Alikhanyan National Science Laboratory, 0036
Yerevan, Armenia.}
\affiliation{$^{2}$ Applied Mathematics Research
Centre, Coventry University, Coventry, CV1 5FB, England, United
Kingdom.}

\begin{abstract}
The exactly solvable spin-$\frac{1}{2}$ Ising-Heisenberg model on
diamond chain has been considered. We have found the exact results
for the magnetization by using recursion relation method. The
existence of the magnetization plateau has been observed at one
third of the saturation magnetization in the antiferromagnetic case.
Some ground-state properties of the model are examined. At low
temperatures, the system has two ferrimagnetic (FRI1 and FRI2)
phases and one paramagnetic (PRM) phase. Lyapunov exponents for the
various values of the exchange parameters and temperatures have been
analyzed. It have also been shown that the maximal Lyapunov exponent
exhibits plateau. Lyapunov exponents exhibit different behavior for
two ferromagnetic phases. We have found the existence of the
supercritical point for the multi-dimensional rational mapping of
the spin-$\frac{1}{2}$ Ising-Heisenberg model on diamond chain for
the first time at absence of the external magnetic field and $T
\rightarrow 0$ in the antiferromagnetic case.
\end{abstract}
\maketitle

\section{Introduction}

The investigation of physical properties of the low-dimensional
quantum spin systems with competing interactions in an external
magnetic field has been a subject of increasingly intense research
interest in the recent decades. The research interest of these
systems has attracted much attention due to the following reasons:
first, they can be solved exactly by using different mathematical
techniques, second, they are realized in the nature, and third,
these systems present rich thermodynamic behavior, such as the
appearance of magnetization plateaus, double peaks structure of the
specific heat and magnetic susceptibility.

One of the interesting low-dimensional quantum spin system is the
frustrated diamond Heisenberg spin-chain. The physical properties of
real materials such as copper mineral
$\mathrm{Cu_{3}(CO_{3})_{2}(OH)_{2}}$, known as natural azurite
(Copper Carbonate Hydroxide) can be well described by using the
quantum antiferromagnetic Heisenberg model on a generalized diamond
chain. The physical properties of the Heisenberg model on diamond
chain have been investigated using different methods they are full
numerical diagonalization and the Lanczos algorithm
\cite{Mikeska,Derzhko,Okamoto}, the decoration-iteration
transformation \cite{Canova,Lisnii,Pereira,Rojas}, the mapping
transformation technique \cite{Strecka}, the density-matrix
renormalization-group (DMRG) and transfer-matrix
renormalization-group (TMRG) techniques \cite{Gu}, Gibbs-Bogoliubov
approach \cite{Ananikian}, cluster approach \cite{Ananikian6}, the
generalized gradient approximations (GGA) \cite{Kang}, the density
functional theory and state-of-the-art numerical many body
calculations \cite{Jeschke}.

Intriguing properties of the azurite made it a good candidate for
studying its properties on the low-dimensional quantum spin systems.
Kikuchi and co-workers \cite{Kikuchi} have experimentally studied
the physical properties of the compound
$\mathrm{Cu_{3}(CO_{3})_{2}(OH)_{2}}$. They have shown that azurite
can be regarded as a model substance of a distorted diamond chain.
The temperature dependence on the magnetic susceptibility and
specific heat shows double peak structure (around 20 and 5 K) on
both magnetic susceptibility and specific heat results. The
existence of the magnetization plateau at one third of the
saturation magnetization has also been experimentally observed in
the magnetization curve.

The aim of the recursion relation method is to cut lattice into
branches and express the partition function of all lattice through
the partition function of branches. This procedure will allow to
derive one- or multi-dimensional mapping for branches of the
partition function. After which the thermodynamic quantities of the
physical system such as magnetization, magnetic susceptibility,
specific heat can be expressed through recursion relation. One and
multi dimensional mapping allows to investigate properties of
different models for example Ising model on Husimi lattice
\cite{Ananikian2, Ananikian3}, zigzag ladder \cite{Hovhannisyan},
triangular lattice \cite{Arekelyan}, two-layer Bethe lattice
\cite{Izmailian,Albayrak}, mixed-spin Ising model on a decorated
Bethe lattice \cite{Strecka2,Albayrak2}, Q-state Potts model on the
Bethe lattice \cite{Ananikian3}, zigzag ladder \cite{Ananikian4},
phase diagrams for both ferromagnetic and antiferromagnetic cases,
multicritical points, for the spin-1 Ising model on the Bethe
lattice \cite{Akheyan,Ananikian5,Avakian,Ananikian1}.

In this paper we have investigated some properties of the
spin-$\frac{1}{2}$ Ising-Heisenberg model on diamond chain by using
dynamical system (recursion relation) approach. Especially we have
investigated magnetic properties of the model and shown the
existence of the magnetization plateau at one third of the
saturation value. The investigation of the ground-state properties
of the model in the $\Delta - h$ plane shows the existence of three
phases in the antiferromagnetic case and two phases in the
ferromagnetic case. Another interesting property of the model has
been found by investigating the behavior of Lyapunov exponent.
Especially we have shown the existence of the plateau in the maximal
Lyapunov exponent curve.

The rest of the paper is organized as follows: In the next section
using the recursion relation method we derive the exact two
dimensional recursion relations for the partition function of the
spin-$\frac{1}{2}$ Ising-Heisenberg model on diamond chain. The
exact results for the magnetization of Ising and Heisenberg spin
sublattices have been derived. We describe the ground-state
properties of the model in $\Delta - h$ plane. In Sec. \ref{Lyap} we
have discussed the behavior of Lyapunov exponent. For the
antiferromagnetic case the maximal Lyapunov exponent for the
multi-dimensional rational mapping is considered and it is shown
that near the magnetization plateaus the maximal Lyapunov exponent
also exhibits plateau structure. The supercritical point at $h=0$
and $T\rightarrow0$ has been found. Finally, section
\ref{Conclusion} contains concluding remarks.

\section{Recursion Relation for the Ising-Heisenberg diamond chain}

Let us consider the spin-$\frac{1}{2}$ Ising-Heisenberg model on
diamond chain with free boundary conditions in the presence of an
external magnetic field. The Hamiltonian operator of the model
is equal to the summation of the plaquette Hamiltonians and can be written as
\begin{align}
\mathcal{H}= &\sum_{i=1}^{N}\mathcal{H}_{i}=
\sum_{i=1}^{N}[J(S_{a,i}^{x}S_{b,i}^{x}+S_{a,i}^{y}S_{b,i}^{y}+\Delta
S_{a,i}^{z}S_{b,i}^{z})+J_{1}\left(S_{a,i}^{z}+S_{b,i}^{z}\right)\left(\mu_{i}^{z}+\mu_{i+1}^{z}\right)\nonumber \\
 & -h_{H}\left(S_{a,i}^{z}+S_{b,i}^{z}\right)-\frac{h_{I}}{2}\left(\mu_{i}^{z}+\mu_{i+1}^{z}\right)],\label{eq:Hamil-1}
\end{align}
where $\mathcal{H}_{i}$ is Hamiltonian of each plaquette,
$S_{a,i}^{\alpha}$, $S_{b,i}^{\alpha}$ ($\alpha=x,y,z$) and
$\mu_{i}^{z}$ represent relevant components of Heisenberg
spin-$\frac{1}{2}$ and Ising spin-$\frac{1}{2}$ operators, the
parameters $J$ and $J_1$ stand for the interaction between the
nearest-neighbouring Heisenberg pairs and the nearest-neighbouring
Ising and Heisenberg spins, respectively and $\Delta$ is the
anisotropy parameter. Hamiltonian (1) also includes longitudinal
external magnetic fields $h_{H}$ and $h_I$ interacting with
Heisenberg and Ising spins. The first summation in Eq. (1) is
corresponding to the interstitial anisotropic Heisenberg spins
coupling ($J$ and $\Delta$), the second summation is corresponding
to the interaction between the nearest Ising and Heisenberg spins
and the last two summations are corresponding to the field
interaction with Ising and Heisenberg spins. In our further
calculations we will consider the case when external magnetic field
is uniform $h_{H}=h_{I}$. It is important to mention the separable
nature of the Ising-type exchange interactions between neighboring
Heisenberg dimers which are caused from the following commutation
rule between different plaquette Hamiltonians:
$[\mathcal{H}_i,\mathcal{H}_j]=0$ for $i \neq j$.

The partition function of the system with Hamiltonian (1) is
\begin{equation}
Z=\sum_{\{\mu_i,S_{a,i},S_{b,i}\}}exp\{-{\beta}{\mathcal{H}}\},\label{Partition}
\end{equation}
where $\beta=(k_{B}T)^{-1}$, $k_{B}$ is Boltzmann constant
(hereafter we consider $k_B=1$) and T is the absolute temperature.
By cutting diamond chain at $S_{a,0}$ and $S_{b,0}$ points (central plaquette) into two
branches (we denote these branches $g_n({S_{a,0}},S_{b,0})$ see Fig.
1) the exact recursion relation for the partition function can be
derived. After this procedure the partition function can be written
as
\begin{equation}
Z=\sum_{\{S_{a,0},S_{b,0}\}}e^{-{\beta}[J(S_{a,0}^{x}S_{b,0}^{x}+S_{a,0}^{y}S_{b,0}^{y}+\Delta
S_{a,0}^{z}S_{b,0}^{z})-h(S_{a,0}^{z}+S_{b,0}^{z})]}g^{2}_{n}(S_{a,0},S_{b,0}),\label{eq3}
\end{equation}
where $g^{2}_{n}(S_{a,0},S_{b,0})$ is contribution of both left and
right branches. The sum in Eq. (\ref{eq3}) goes over all possible
combinations of Heisenberg spins $S_{a,0}$ and $S_{b,0}$. Putting
into Eq. (\ref{eq3}) eigenvalues of the operator
$exp\{-{\beta}[J(S_{a,0}^{x}S_{b,0}^{x}+S_{a,0}^{y}S_{b,0}^{y}+\Delta
S_{a,0}^{z}S_{b,0}^{z})-h(S_{a,0}^{z}+S_{b,0}^{z})]\}$ we can get
the partition function expressed through $g_{n}(S_{a,0},S_{b,0})$
\begin{equation}
Z=e^{-\frac{J\Delta}{4T}+\frac{h}{T}}g^{2}_{n}(\uparrow\uparrow)
+e^{-\frac{J}{2T}+\frac{J\Delta}{4T}}g^{2}_{n}(\frac{\uparrow\downarrow+\downarrow\uparrow}{\sqrt{2}})
+e^{\frac{J}{2T}+\frac{J\Delta}{4T}}g^{2}_{n}(\frac{\uparrow\downarrow-\downarrow\uparrow}{\sqrt{2}})
+e^{-\frac{J\Delta}{4T}-\frac{h}{T}}g^{2}_{n}(\downarrow\downarrow),\label{eq4}
\end{equation}
where by $\uparrow$ (up) and $\downarrow$ (down) we denote directions of
Heisenberg spins. To find recursion relations for the model we
need to find relations between $g_{n}(S_{a,0},S_{b,0})$ and
$g_{n-1}(S_{a,1},S_{b,1})$.
\begin{align}
g_n(S_{a,0}, S_{b,0})=&\sum_{\{\mu_1,S_{a,1},S_{b,1}\}}
e^{-{\beta}[J(S_{a,1}^{x}S_{b,1}^{x}+S_{a,1}^{y}S_{b,1}^{y}+\Delta
S_{a,1}^{z}S_{b,1}^{z})
+J_{1}(S_{a,0}^{z}+S_{b,0}^{z})\mu_{1}^{z}
+J_{1}\mu_{1}^{z}(S_{a,1}^{z}+S_{b,1}^{z})
-h(S_{a,1}^{z}+S_{b,1}^{z}+\mu_{1}^{z})]}
\\ \nonumber
&\ast g_{n-1}(S_{a,1},S_{b,1})
.\label{re}
\end{align}

\begin{figure}
\includegraphics[scale=0.7]{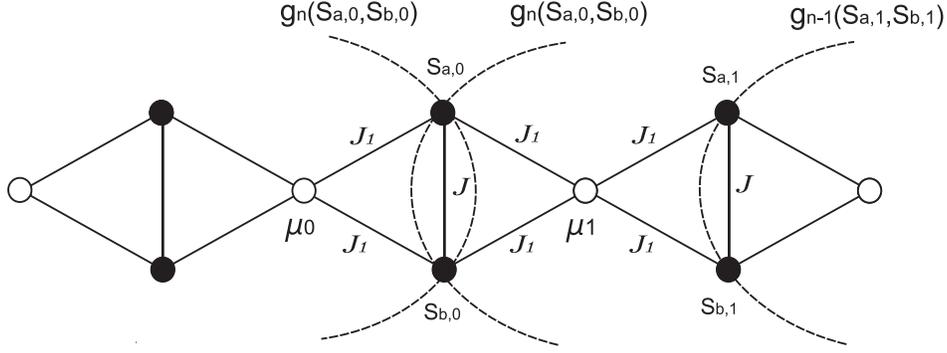}\caption{\label{fig:diamond}The procedure for derivation of the diamond chain.}
\end{figure}

Inserting into Eq. (5) eigenvalues of the operator
$exp\{-{\beta}[J(S_{a,1}^{x}S_{b,1}^{x}+S_{a,1}^{y}S_{b,1}^{y}+\Delta
S_{a,1}^{z}S_{b,1}^{z}) +J_{1}(S_{a,0}^{z}+S_{b,0}^{z})\mu_{1}^{z}
+J_{1}\mu_{1}^{z}(S_{a,1}^{z}+S_{b,1}^{z})
-h(S_{a,1}^{z}+S_{b,1}^{z}+\mu_{1}^{z})]\}$ we can express
$g_{n}(S_{a,0},S_{b,0})$ through $g_{n-1}(S_{a,1},S_{b,1})$

\begin{align}
g_n(\uparrow\uparrow)=&(e^{-\frac{J\Delta}{4T}-\frac{J_1}{T}+\frac{3h}{2T}}+e^{-\frac{J\Delta}{4T}+\frac{J_1}{T}+\frac{h}{2T}})g_{n-1}(\uparrow\uparrow)
+(e^{-\frac{J}{2T}+\frac{J\Delta}{4T}-\frac{J_1}{2T}+\frac{h}{2T}}+e^{-\frac{J}{2T}+\frac{J\Delta}{4T}+\frac{J_1}{2T}-\frac{h}{2T}})g_{n-1}(\frac{\uparrow\downarrow+\downarrow\uparrow}{\sqrt{2}})
\\ \nonumber
&+(e^{\frac{J}{2T}+\frac{J\Delta}{4T}-\frac{J_1}{2T}+\frac{h}{2T}}+e^{\frac{J}{2T}+\frac{J\Delta}{4T}+\frac{J_1}{2T}-\frac{h}{2T}})g_{n-1}(\frac{\uparrow\downarrow-\downarrow\uparrow}{\sqrt{2}})
+e^{-\frac{J\Delta}{4T}-\frac{h}{2T}}+e^{-\frac{J\Delta}{4T}-\frac{3h}{2T}})g_{n-1}(\downarrow\downarrow),
\\ \nonumber
g_n(\frac{\uparrow\downarrow+\downarrow\uparrow}{\sqrt{2}})=&(e^{-\frac{J\Delta}{4T}-\frac{J_1}{2T}+\frac{3h}{2T}}+e^{-\frac{J\Delta}{4T}+\frac{J_1}{2T}+\frac{h}{2T}})g_{n-1}(\uparrow\uparrow)
+(e^{-\frac{J}{2T}+\frac{J\Delta}{4T}+\frac{h}{2T}}+e^{-\frac{J}{2T}+\frac{J\Delta}{4T}-\frac{h}{2T}})g_{n-1}(\frac{\uparrow\downarrow+\downarrow\uparrow}{\sqrt{2}})
\\ \nonumber
&+(e^{\frac{J}{2T}+\frac{J\Delta}{4T}+\frac{h}{2T}}+e^{\frac{J}{2T}+\frac{J\Delta}{4T}-\frac{h}{2T}})g_{n-1}(\frac{\uparrow\downarrow-\downarrow\uparrow}{\sqrt{2}})
+(e^{-\frac{J\Delta}{4T}+\frac{J_1}{2T}-\frac{h}{2T}}+e^{-\frac{J\Delta}{4T}-\frac{J_1}{2T}-\frac{3h}{2T}})g_{n-1}(\downarrow\downarrow),
\\ \nonumber
g_n(\frac{\uparrow\downarrow-\downarrow\uparrow}{\sqrt{2}})=&(e^{-\frac{J\Delta}{4T}-\frac{J_1}{2T}+\frac{3h}{2T}}+e^{-\frac{J\Delta}{4T}+\frac{J_1}{2T}+\frac{h}{2T}})g_{n-1}(\uparrow\uparrow)
+(e^{-\frac{J}{2T}+\frac{J\Delta}{4T}+\frac{h}{2T}}+e^{-\frac{J}{2T}+\frac{J\Delta}{4T}-\frac{h}{2T}})g_{n-1}(\frac{\uparrow\downarrow+\downarrow\uparrow}{\sqrt{2}})
\\ \nonumber
&+(e^{\frac{J}{2T}+\frac{J\Delta}{4T}+\frac{h}{2T}}+e^{\frac{J}{2T}+\frac{J\Delta}{4T}-\frac{h}{2T}})g_{n-1}(\frac{\uparrow\downarrow-\downarrow\uparrow}{\sqrt{2}})
+(e^{-\frac{J\Delta}{4T}+\frac{J_1}{2T}-\frac{h}{2T}}+e^{-\frac{J\Delta}{4T}-\frac{J_1}{2T}-\frac{3h}{2T}})g_{n-1}(\downarrow\downarrow),
\\ \nonumber
g_n(\downarrow\downarrow)=&(e^{-\frac{J\Delta}{4T}+\frac{3h}{2T}}+e^{-\frac{J\Delta}{4T}+\frac{h}{2T}})g_{n-1}(\uparrow\uparrow)
+(e^{-\frac{J}{2T}+\frac{J\Delta}{4T}+\frac{J_1}{2T}+\frac{h}{2T}}+e^{-\frac{J}{2T}+\frac{J\Delta}{4T}-\frac{J_1}{2T}-\frac{h}{2T}})g_{n-1}(\frac{\uparrow\downarrow+\downarrow\uparrow}{\sqrt{2}})
\\ \nonumber
&+(e^{\frac{J}{2T}+\frac{J\Delta}{4T}+\frac{J_1}{2T}+\frac{h}{2T}}+e^{\frac{J}{2T}+\frac{J\Delta}{4T}-\frac{J_1}{2T}-\frac{h}{2T}})g_{n-1}(\frac{\uparrow\downarrow-\downarrow\uparrow}{\sqrt{2}})
+(e^{-\frac{J\Delta}{4T}+\frac{J_1}{T}-\frac{h}{2T}}+e^{-\frac{J\Delta}{4T}-\frac{J_1}{T}-\frac{3h}{2T}})g_{n-1}(\downarrow\downarrow).\label{rec}
\end{align}
As it can be seen from relations (6)
$g_n(\frac{\uparrow\downarrow+\downarrow\uparrow}{\sqrt{2}})=g_n(\frac{\uparrow\downarrow-\downarrow\uparrow}{\sqrt{2}})$
hence our recursion relation will be two-dimensional rational
mapping. By introducing the following notations
\begin{eqnarray}
x_n=\frac{g_n(\uparrow\uparrow)}{g_n(\frac{\uparrow\downarrow+\downarrow\uparrow}{\sqrt{2}})},
 \\ \nonumber
y_n=\frac{g_n(\downarrow\downarrow)}{g_n(\frac{\uparrow\downarrow+\downarrow\uparrow}{\sqrt{2}})},
\label{eq:rec1}
\end{eqnarray}
we can get two-dimensional recursion relation for the partition
function
\begin{align}
x_n=&[(e^{-\frac{J\Delta}{4T}-\frac{J_1}{T}+\frac{3h}{2T}}
+e^{-\frac{J\Delta}{4T}+\frac{J_1}{T}+\frac{h}{2T}})x_{n-1}
+e^{-\frac{J}{2T}+\frac{J\Delta}{4T}-\frac{J_1}{2T}+\frac{h}{2T}}
+e^{-\frac{J}{2T}+\frac{J\Delta}{4T}+\frac{J_1}{2T}-\frac{h}{2T}}
\\ \nonumber
&+e^{\frac{J}{2T}+\frac{J\Delta}{4T}-\frac{J_1}{2T}+\frac{h}{2T}}
+e^{\frac{J}{2T}+\frac{J\Delta}{4T}+\frac{J_1}{2T}-\frac{h}{2T}}
+(e^{-\frac{J\Delta}{4T}-\frac{h}{2T}}
+e^{-\frac{J\Delta}{4T}-\frac{3h}{2T}})y_{n-1}]
\\ \nonumber
&/[(e^{-\frac{J\Delta}{4T}-\frac{J_1}{2T}+\frac{3h}{2T}}
+e^{-\frac{J\Delta}{4T}+\frac{J_1}{2T}+\frac{h}{2T}})x_{n-1}
+e^{-\frac{J}{2T}+\frac{J\Delta}{4T}+\frac{h}{2T}}
+e^{-\frac{J}{2T}+\frac{J\Delta}{4T}-\frac{h}{2T}}
\\ \nonumber
&+e^{\frac{J}{2T}+\frac{J\Delta}{4T}+\frac{h}{2T}}
+e^{\frac{J}{2T}+\frac{J\Delta}{4T}-\frac{h}{2T}}
+(e^{-\frac{J\Delta}{4T}+\frac{J_1}{2T}-\frac{h}{2T}}
+e^{-\frac{J\Delta}{4T}-\frac{J_1}{2T}-\frac{3h}{2T}})y_{n-1}]
\\ \nonumber
y_n=&[(e^{-\frac{J\Delta}{4T}+\frac{3h}{2T}}
+e^{-\frac{J\Delta}{4T}+\frac{h}{2T}})x_{n-1}
+e^{-\frac{J}{2T}+\frac{J\Delta}{4T}+\frac{J_1}{2T}+\frac{h}{2T}}
+e^{-\frac{J}{2T}+\frac{J\Delta}{4T}-\frac{J_1}{2T}-\frac{h}{2T}}
\\ \nonumber
&+e^{\frac{J}{2T}+\frac{J\Delta}{4T}+\frac{J_1}{2T}+\frac{h}{2T}}
+e^{\frac{J}{2T}+\frac{J\Delta}{4T}-\frac{J_1}{2T}-\frac{h}{2T}}
+(e^{-\frac{J\Delta}{4T}+\frac{J_1}{T}-\frac{h}{2T}}
+e^{-\frac{J\Delta}{4T}-\frac{J_1}{T}-\frac{3h}{2T}})y_{n-1}]
\\ \nonumber
&/[(e^{-\frac{J\Delta}{4T}-\frac{J_1}{2T}+\frac{3h}{2T}}
+e^{-\frac{J\Delta}{4T}+\frac{J_1}{2T}+\frac{h}{2T}})x_{n-1}
+e^{-\frac{J}{2T}+\frac{J\Delta}{4T}+\frac{h}{2T}}
+e^{-\frac{J}{2T}+\frac{J\Delta}{4T}-\frac{h}{2T}}
\\ \nonumber
&+e^{\frac{J}{2T}+\frac{J\Delta}{4T}+\frac{h}{2T}}
+e^{\frac{J}{2T}+\frac{J\Delta}{4T}-\frac{h}{2T}}
+(e^{-\frac{J\Delta}{4T}+\frac{J_1}{2T}-\frac{h}{2T}}
+e^{-\frac{J\Delta}{4T}-\frac{J_1}{2T}-\frac{3h}{2T}})y_{n-1}].
\label{recursion}
\end{align}

Recursion relation (8) plays a crucial role in our further
investigation because the thermodynamic quantities like
magnetization can be expressed through two-dimensional rational
mapping. Magnetization for the sublattice of Heisenberg spins can be
found using the following formula

\begin{eqnarray}
m_H&=&\frac{<S_{a,i}^{z}+S_{b,i}^{z}>}{2}=\frac{<S_{a,0}^{z}+S_{b,0}^{z}>}{2}
\\ \nonumber
&=&\frac{\sum_{\{S_{a,0},S_{b,0}\}}(S_{a,0}^{z}+S_{b,0}^{z})
e^{-{\beta}{[J(S_{a,0}^{x}S_{b,0}^{x}+S_{a,0}^{y}S_{b,0}^{y}+\Delta
S_{a,0}^{z}S_{b,0}^{z})-h\left(S_{a,0}^{z}+S_{b,0}^{z}\right)]}}g^{2}_{n}(S_{a,0},S_{b,0})}{2Z}.
\label{HeisenbergMag}
\end{eqnarray}

In Eq. (9) the sum goes over all possible combinations of $S_{a,0}$
and $S_{b,0}$. Putting into Eq. (9) expression for the partition
function and taking into account the notation (7) we can express
magnetization for the sublattice of Heisenberg spins through
recursion relations which can be written as

\begin{figure}[!h]
\begin{center}
\begin{tabular}{cccc}
{\small (a)}& {\small (b)}\\
\includegraphics[width=8cm]{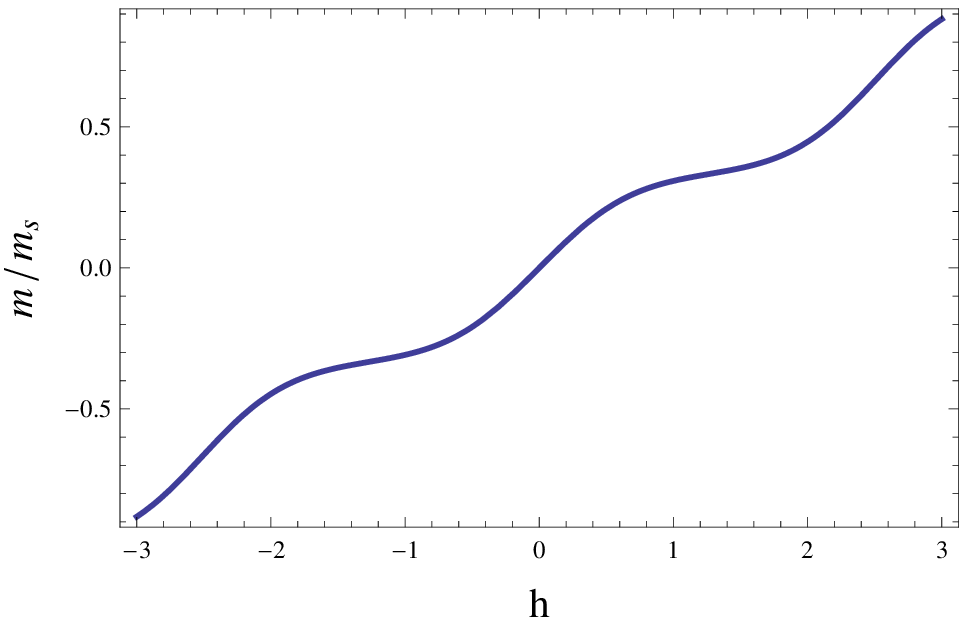} &
\includegraphics[width=8cm]{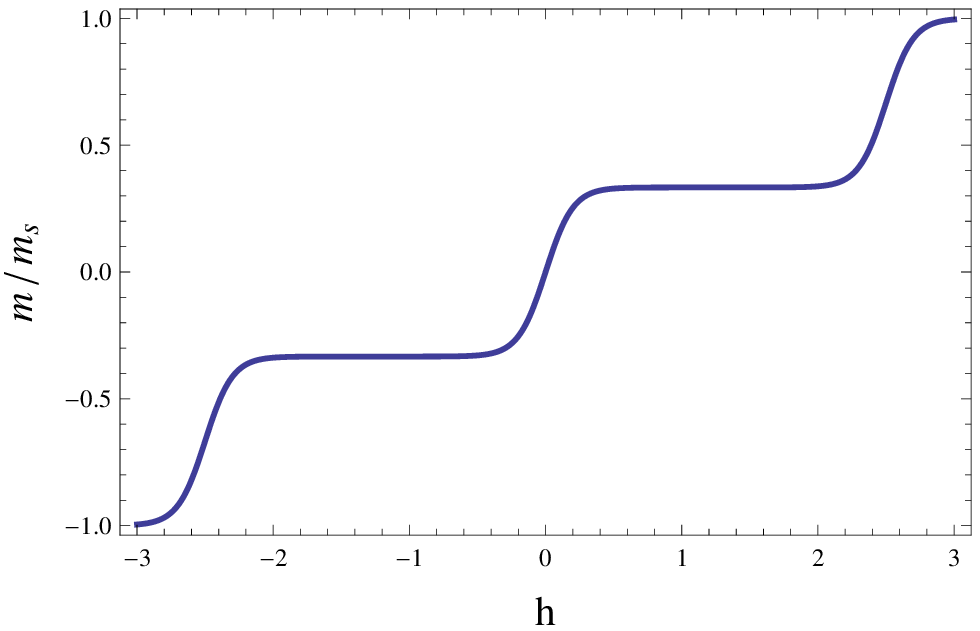}&
\\ \nonumber
{\small (c)}& {\small (d)}\\
\includegraphics[width=8cm]{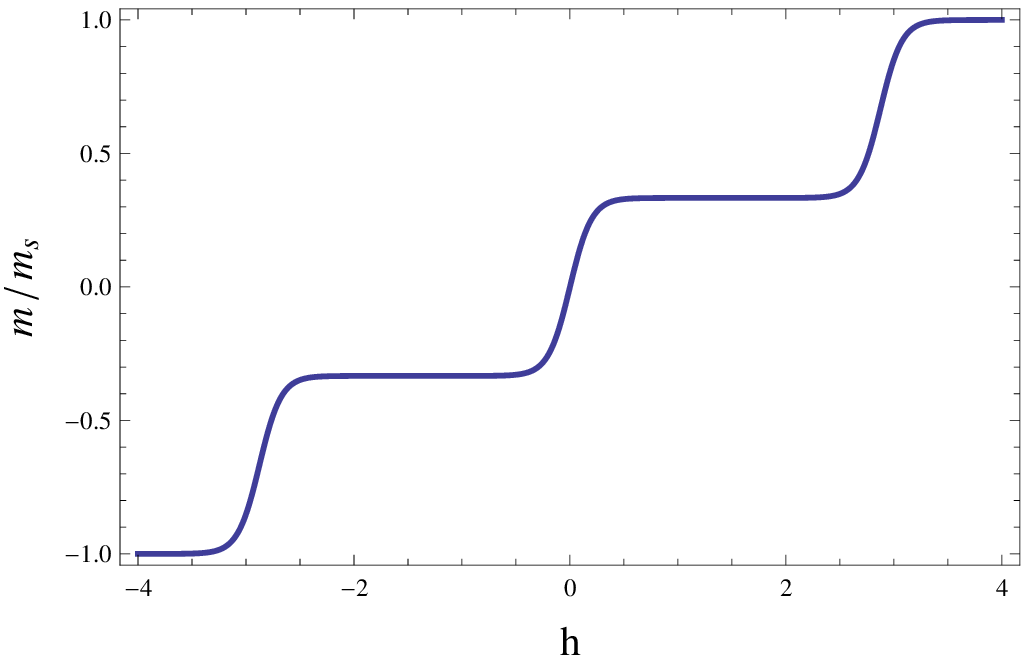} &
\includegraphics[width=8cm]{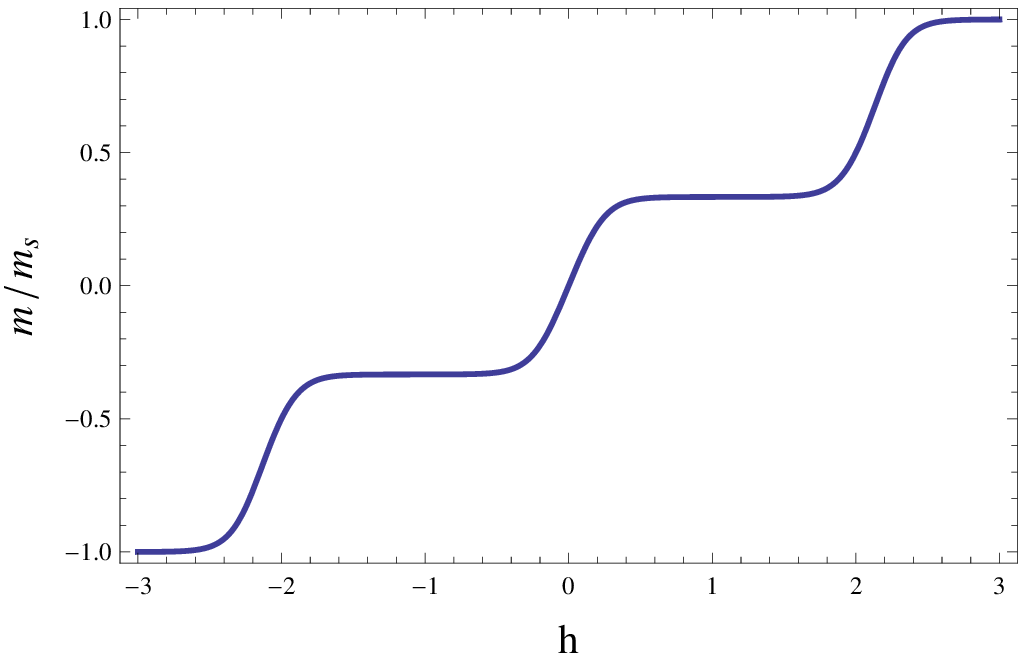}&
\\ \nonumber
{\small (e)}& {\small (f)} \\
\includegraphics[width=8cm]{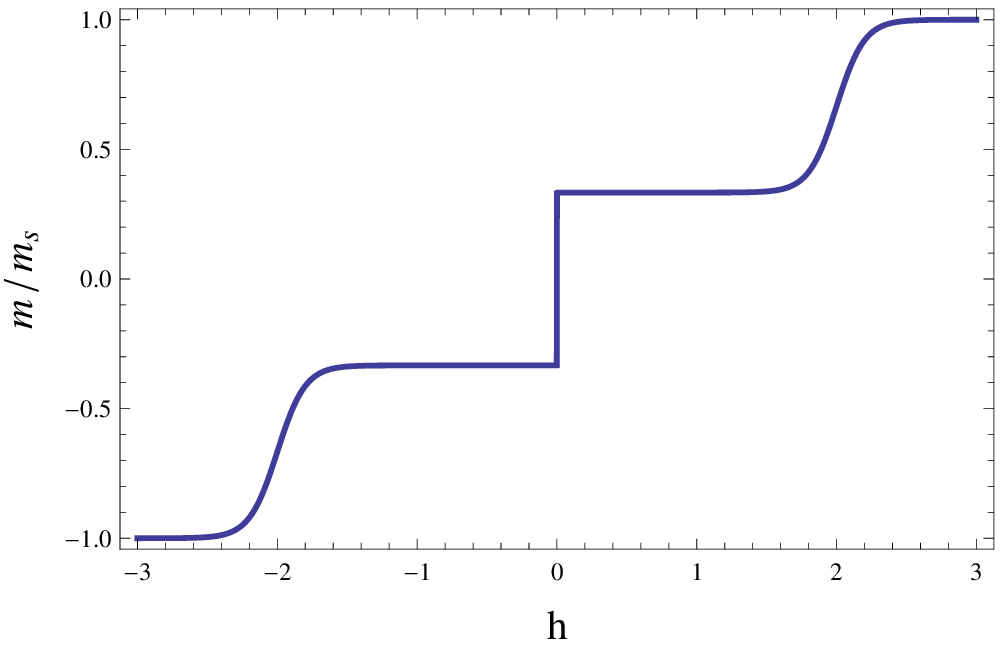} &
\includegraphics[width=8cm]{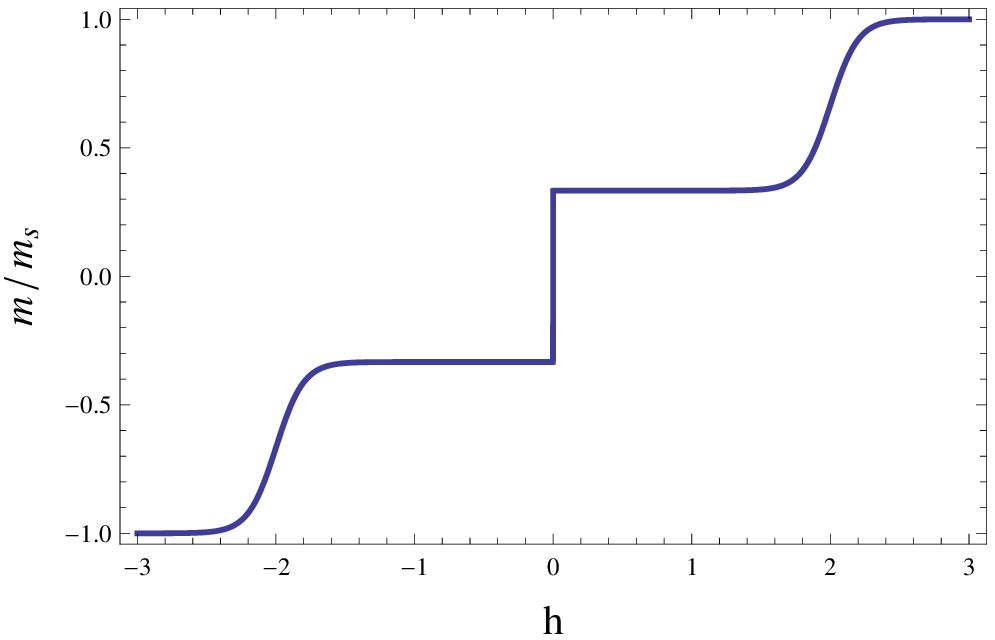}\\
\end{tabular}
\caption { \small{The field dependence of the total magnetization
with respect to its saturation value at exchange parameters $J=1.5$
and $J_1=1$: (a) $T=0.3$, $\Delta=1$; (b) $T=0.1$, $\Delta=1$; (c)
$T=0.1$, $\Delta=1.5$; (d) $T=0.1$, $\Delta=0.5$; (e) $T=0.1$,
$\Delta=-1$; (f) $T=0.1$, $\Delta=-1.5$.}} \label{mag1}
\end{center}
\end{figure}

\begin{figure}[!h]
\begin{center}
\begin{tabular}{cc}
{\small (a)}& {\small (b)}\\
\includegraphics[width=7cm]{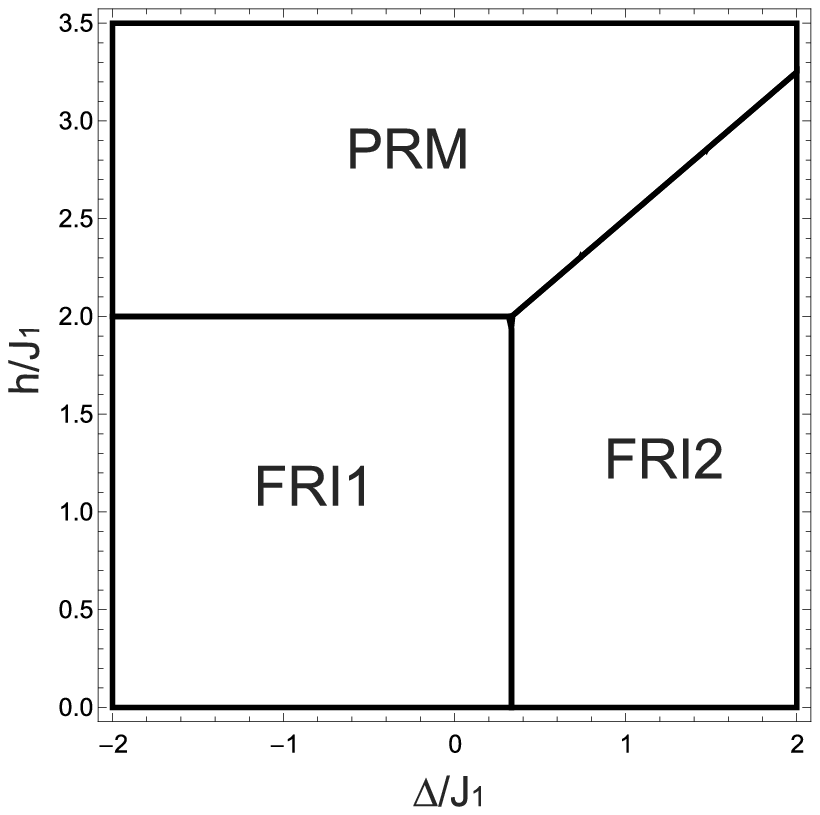} &
\includegraphics[width=7cm]{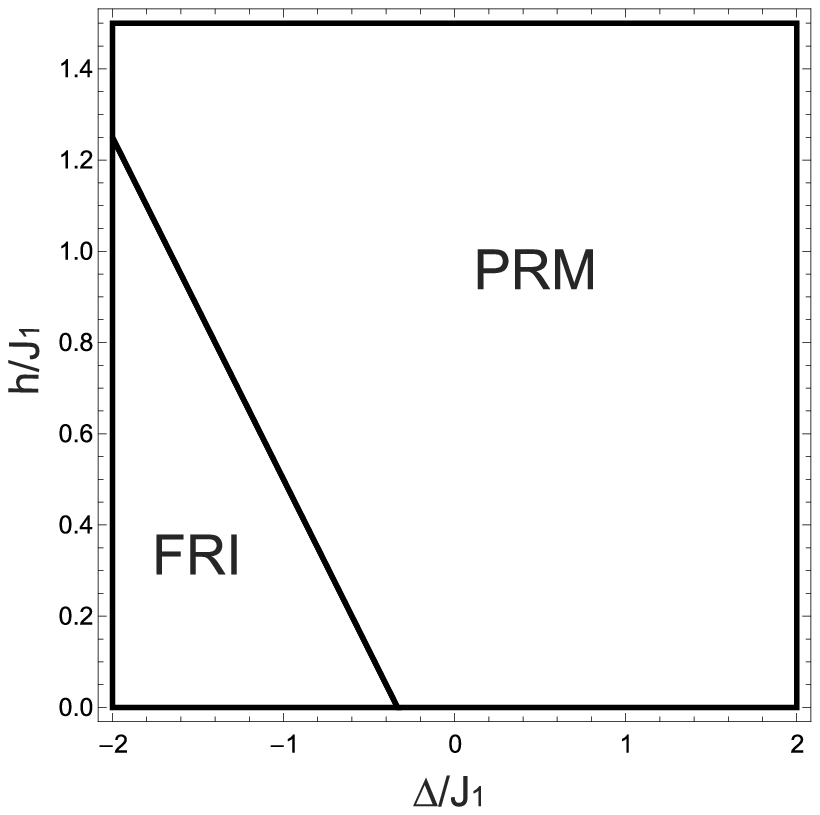}\\
\end{tabular}
\caption { \small{Ground-state phase diagram in the $\Delta-h$ plane
for (a) antiferromagnetic case $J=1.5$, $J_1=1$ (b) ferromagnetic
case $J=-1.5$, $J_1=-1$.}} \label{Phase}
\end{center}
\end{figure}

\begin{eqnarray}
m_H=\frac{e^{-\frac{J\Delta}{4T}+\frac{h}{T}}x^{2}_{n}-e^{-\frac{J\Delta}{4T}-
\frac{h}{T}}y^{2}_{n}}{2(e^{-\frac{J\Delta}{4T}+\frac{h}{T}}x^{2}_{n}+e^{-\frac{J}{2T}+\frac{J\Delta}{4T}}
+e^{\frac{J}{2T}+\frac{J\Delta}{4T}}+e^{-\frac{J\Delta}{4T}-\frac{h}{T}}y^{2}_{n})}.
\label{Mag1}
\end{eqnarray}

In the same way we can find magnetization for the sublattice of Ising spins.
\begin{eqnarray}
m_I=<\mu_i>=<\mu_1>=\frac{\sum_{\{\mu_i,S_{a,0},S_{b,0}\}}\mu_1 e^{-{\beta}{[J(S_{a,0}^{x}S_{b,0}^{x}+S_{a,0}^{y}S_{b,0}^{y}+\Delta
S_{a,0}^{z}S_{b,0}^{z})-h\left(S_{a,0}^{z}+S_{b,0}^{z}\right)]}}g^{2}_{n}(S_{a,0},S_{b,0})}{Z}.
\label{Mag2}
\end{eqnarray}

In this expression $\mu_1$ is a part of right branch of
$g_{n}(S_{a,0},S_{b,0})$. So to find magnetization for the
sublattice of Ising spins we need to express
$g_{n}(S_{a,0},S_{b,0})$ through $g_{n-1}(S_{a,1},S_{b,1})$. It is
important to mention that this procedure also should be done for the
partition function. After this procedure the expression for the
magnetization of the sublattice of Ising spins can be expressed through recursion relation
(8):
\begin{align}
m_I=&[(f_{1}(x_{n-1},y_{n-1})-f_{2}(x_{n-1},y_{n-1}))e^{-\frac{J\Delta}{4T}+\frac{h}{T}}x_n+
(f_{3}(x_{n-1},y_{n-1})-f_{4}(x_{n-1},y_{n-1}))(e^{-\frac{J}{2T}+\frac{J\Delta}{4T}}+e^{\frac{J}{2T}+\frac{J\Delta}{4T}})
\\ \nonumber
&+(f_{5}(x_{n-1},y_{n-1})-f_{6}(x_{n-1},y_{n-1}))e^{-\frac{J\Delta}{4T}-\frac{h}{T}}y_n]
/[(f_{1}(x_{n-1},y_{n-1})+f_{2}(x_{n-1},y_{n-1}))e^{-\frac{J\Delta}{4T}+\frac{h}{T}}x_n
\\ \nonumber
&+(f_{3}(x_{n-1},y_{n-1})+f_{4}(x_{n-1},y_{n-1}))(e^{-\frac{J}{2T}+\frac{J\Delta}{4T}}+e^{\frac{J}{2T}+\frac{J\Delta}{4T}})
+(f_{5}(x_{n-1},y_{n-1})+f_{6}(x_{n-1},y_{n-1}))e^{-\frac{J\Delta}{4T}-\frac{h}{T}}y_n]
,\label{MagnetizationIsing}
\end{align}
where

\begin{eqnarray}
f_{1}(x,y)&=&e^{-\frac{J\Delta}{4T}-\frac{J_1}{T}+\frac{3h}{2T}}x
+e^{-\frac{J}{2T}+\frac{J\Delta}{4T}-\frac{J_1}{2T}+\frac{h}{2T}}
+e^{\frac{J}{2T}+\frac{J\Delta}{4T}-\frac{J_1}{2T}+\frac{h}{2T}}
+e^{-\frac{J\Delta}{4T}-\frac{h}{2T}}y,
\\ \nonumber
f_{2}(x,y)&=&e^{-\frac{J\Delta}{4T}+\frac{J_1}{T}+\frac{h}{2T}}x
+e^{-\frac{J}{2T}+\frac{J\Delta}{4T}+\frac{J_1}{2T}-\frac{h}{2T}}
+e^{\frac{J}{2T}+\frac{J\Delta}{4T}+\frac{J_1}{2T}-\frac{h}{2T}}
+e^{-\frac{J\Delta}{4T}-\frac{3h}{2T}}y,
\\ \nonumber
f_{3}(x,y)&=&e^{-\frac{J\Delta}{4T}-\frac{J_1}{2T}+\frac{3h}{2T}}x
+e^{-\frac{J}{2T}+\frac{J\Delta}{4T}+\frac{h}{2T}}
+e^{\frac{J}{2T}+\frac{J\Delta}{4T}+\frac{h}{2T}}
+e^{-\frac{J\Delta}{4T}+\frac{J_1}{2T}-\frac{h}{2T}}y,
\\ \nonumber
f_{4}(x,y)&=&e^{-\frac{J\Delta}{4T}+\frac{J_1}{2T}+\frac{h}{2T}}x
+e^{-\frac{J}{2T}+\frac{J\Delta}{4T}-\frac{h}{2T}}
+e^{\frac{J}{2T}+\frac{J\Delta}{4T}-\frac{h}{2T}}
+e^{-\frac{J\Delta}{4T}-\frac{J_1}{2T}-\frac{3h}{2T}}y,
\\ \nonumber
f_{5}(x,y)&=&e^{-\frac{J\Delta}{4T}+\frac{3h}{2T}}x
+e^{-\frac{J}{2T}+\frac{J\Delta}{4T}+\frac{J_1}{2T}+\frac{h}{2T}}
+e^{\frac{J}{2T}+\frac{J\Delta}{4T}+\frac{J_1}{2T}+\frac{h}{2T}}
+e^{-\frac{J\Delta}{4T}+\frac{J_1}{T}-\frac{h}{2T}}y,
\\ \nonumber
f_{6}(x,y)&=&e^{-\frac{J\Delta}{4T}+\frac{h}{2T}}x
+e^{-\frac{J}{2T}+\frac{J\Delta}{4T}-\frac{J_1}{2T}-\frac{h}{2T}}
+e^{\frac{J}{2T}+\frac{J\Delta}{4T}-\frac{J_1}{2T}-\frac{h}{2T}}
+e^{-\frac{J\Delta}{4T}-\frac{J_1}{T}-\frac{3h}{2T}}y.
\label{IsingMag}
\end{eqnarray}

Expressions (10) and (12) will let us calculate the total
single-site magnetization of the spin-$\frac{1}{2}$ Ising-Heisenberg
model on diamond chain which can be written as
\begin{eqnarray}
m=\frac{m_{I}+2m_{H}}{3}.
\end{eqnarray}
Figure \ref{mag1} shows the field behavior of the total
magnetization for antiferromagnetic case at the fixed values of
interaction constants $J=1.5$ and $J_1=1$, anisotropy parameter
$\Delta=1$ and different values of the absolute temperature ($T$).
At high temperatures the magnetization curve has a monotone
structure (Fig. \ref{mag1} (a)). At lower temperatures the plateau
of magnetization at one third is arising in magnetization curve
(Fig. \ref{mag1} (b)). Other plots of the magnetization curves for
the different values of the anisotropy parameter $\Delta$ are
displayed in Fig. \ref{mag1} ((c), (d), (e), (f)). Figures 2 (b),
(c) and (d) show that the larger positive values of the anisotropy
parameter correspond to the larger width of the magnetization
plateau for the fixed value of the absolute temperature. While for
the negative values of anisotropy parameter magnetization curves
remain the same (Fig. \ref{mag1} (e), (f)). As it can be seen from
the figures recursion relation method results are good agrement with
other methods results such as the decoration-iteration
transformation method (for example see [4] figure 3).

Let us research the ground state of the  spin-$\frac{1}{2}$
Ising-Heisenberg model on diamond chain via $\Delta$ and $h$ for the
antiferromagnetic ($J=1.5$, $J_1=1$) and ferromagnetic ($J=-1.5$,
$J_1=-1$) models. Depending on the value of ratio
$\frac{\Delta}{J_{1}}$ and the magnetic field measured in unites of
$J_{1}$, the system exhibits two ferrimagnetic (FRI1 and FRI2) and
one paramagnetic (PRM) ground-state phases (Fig. 3(a)) for the
antiferromagnetic case. Phases FRI1, FRI2 and PRM correspond to the
following values of Ising and Heisenberg spins sublattice
magnetization:
\begin{eqnarray}
FRI1: m_{I}&=&-0.5, m_{H}=0.5,
\\ \nonumber
FRI2: m_{I}&=&0.5, m_{H}=0,
\\ \nonumber
PRM: m_{I}&=&0.5, m_{H}=0.5. \label{Phases}
\end{eqnarray}
Analytically it can be shown that for the fixed values of exchange
parameters $J=1.5$ and $J_1=1$ phase transition from FRI1 to FRI2
takes place at $\Delta=\frac{1}{3}$. Now let us compare the
displayed magnetization curves (Fig. \ref{mag1}) with the
ground-state phase diagram shown in Fig. 3(a). As it is already
mentioned in FRI2 phase the larger positive values of the anisotropy
parameter ($\Delta>\frac{1}{3}$) correspond to the larger width of
the magnetization plateau see Fig. 3 (b), (c) and (d). In FRI1 phase
for the fixed values of interaction constants and the absolute
temperature the behavior of the magnetization curve remains the same
(Fig. \ref{mag1} (e), (f)).

For the ferromagnetic case there are two phases in the phase
diagram; the ferrimagnetic (FRI) and paramagnetic (PRM) (Fig. 3(b)).
Phases FRI and PRM correspond to the following values of Ising and
Heisenberg spins sublattice magnetization:
\begin{eqnarray}
FRI: m_{I}&=&0.5, m_{H}=0,
\\ \nonumber
PRM: m_{I}&=&0.5, m_{H}=0.5. \label{Phases2}
\end{eqnarray}
It can be analytically shown that FRI phase ends on
$\Delta=-\frac{1}{3}$ at absence of an external magnetic field.

\section{Lyapunov Exponent and superstable point \label{Lyap}}

In this section we will focus on the thermodynamical equilibrium
description of the spin-$\frac{1}{2}$ Ising- Heisenberg model on a
diamond chain, by studying infinite-size systems. Lyapunov exponents
near the magnetization plateau of the antiferromagnetic model are
interesting to calculate on a diamond chain. It is shown that the
behavior of the maximal Lyapunov exponent via magnetic field of
multi-dimensional rational mapping has a plateau and coincides with
magnetization one on one-dimensional kagome chain at low
temperatures \cite{Hrach}. It was obtained that the maximal Lyapunov
exponent had a negative vanishing plateau.

The following values of Lyapunov exponent can be observed during the investigation.

1. $\lambda<0$. Negative Lyapunov exponents show that the system is
dissipative or non-conservative. The systems with more negative
values of Lyapunov exponent are more stabile. If $\lambda=-\infty$
means that we have superstable fixed and superstable periodic
points.

2. $\lambda=0$ corresponding to neutral fixed point. Zero values of
Lyapunov exponents are characteristic for conservative systems. At
this value of Lyapunov exponents the second-order phase transition
takes place.

3. $\lambda>0$ corresponding to unstable and chaotic systems.
The systems with positive Lyapunov exponents have chaotic behavior.

In general for the mapping $x_n=f(x_{n-1})$ Lyapunov exponent
$\lambda(x)$ characterizes the exponential divergence of two nearby
points after $n$ iterations. Lyapunov exponent may be expressed as a
limit of mapping stability as
\cite{Schuster,Oseledec,Eckmann,Crisanti,Latora,Birol,Keskin}
\begin{eqnarray}
\lambda{(x)}=\lim_{n\rightarrow\infty}{\frac{1}{n}}
\ln\mid{\frac{df^{n}(x)}{dx}}\mid. \label{lyap1}
\end{eqnarray}
In multidimensional case with dimension $n$, exists $n$ of exponents
for various directions in space
\begin{eqnarray}
e^{\lambda_{1}},e^{\lambda_{2}},...e^{\lambda_{n}}=\lim_{n\rightarrow\infty}(eigenvalues\ of\ the\ product \prod_{i=0}^{n-1}{J(\overrightarrow{x_{i}})})^{\frac{1}{n}},
\end{eqnarray}
where
$J(\overrightarrow{x})=(\frac{\partial{G_{i}}}{\partial{x_{j}}})$ is the Jacobian of the mapping
$\overrightarrow{x}_{n+1}=G({\overrightarrow{x}_{n}})$. For two dimensional mapping (8)
we can receive the following expression of Lyapunov exponents
\begin{eqnarray}
\lambda_{1},\lambda_{2}=\lim_{n\rightarrow\infty}{\frac{1}{n}}ln({eigenvalues\ of\ the\ product \prod_{i=0}^{n-1}{J(x_{i},y_{i})}})
\label{lyapunov}
\end{eqnarray}

\begin{figure}[!h]
\begin{center}
\begin{tabular}{cc}
{\small (a)}& {\small (b)}\\
\includegraphics[width=8cm]{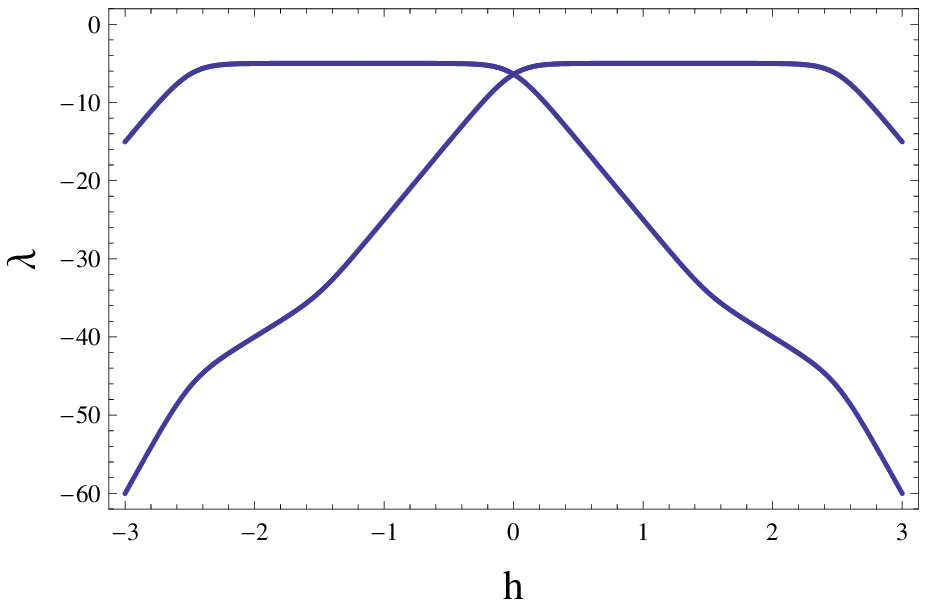} &
\includegraphics[width=8cm]{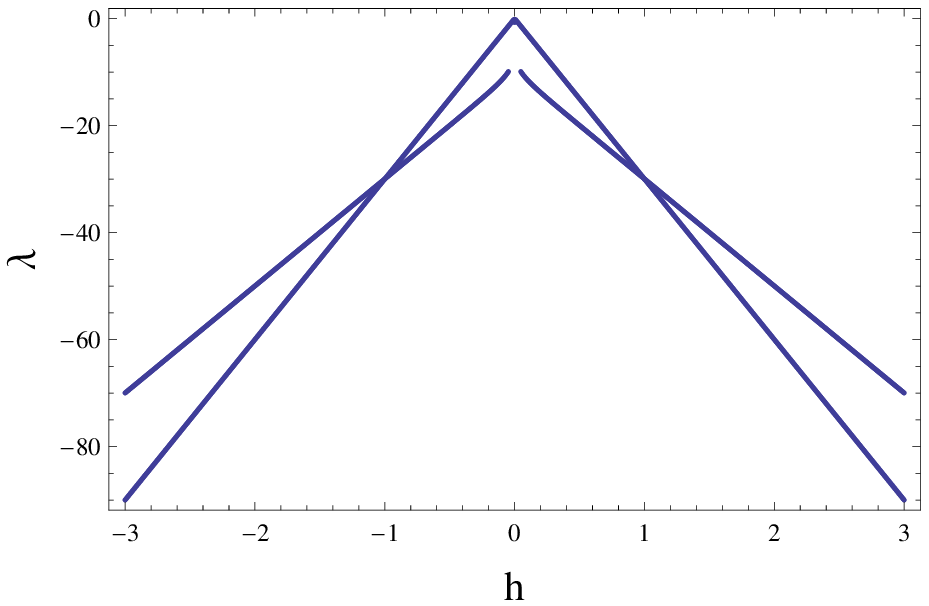}\\
\end{tabular}
\caption { \small{Plot of Lyapunov exponents for spin $\frac{1}{2}$
Ising-Heisenberg model on diamond chain at exchange parameters (a)
the antiferromagnetic case at $J=1.5$, $J_1=1$, $\Delta=1$ and
temperature $T=0.1$ (b) the ferromagnetic case at $J=-1.5$,
$J_1=-1$, $\Delta=1$ and temperature $T=0.1$.}} \label{lyap1}
\end{center}
\end{figure}

\begin{figure}[!h]
\begin{center}
\begin{tabular}{cc}
{\small (a)}& {\small (b)}\\
\includegraphics[width=8cm]{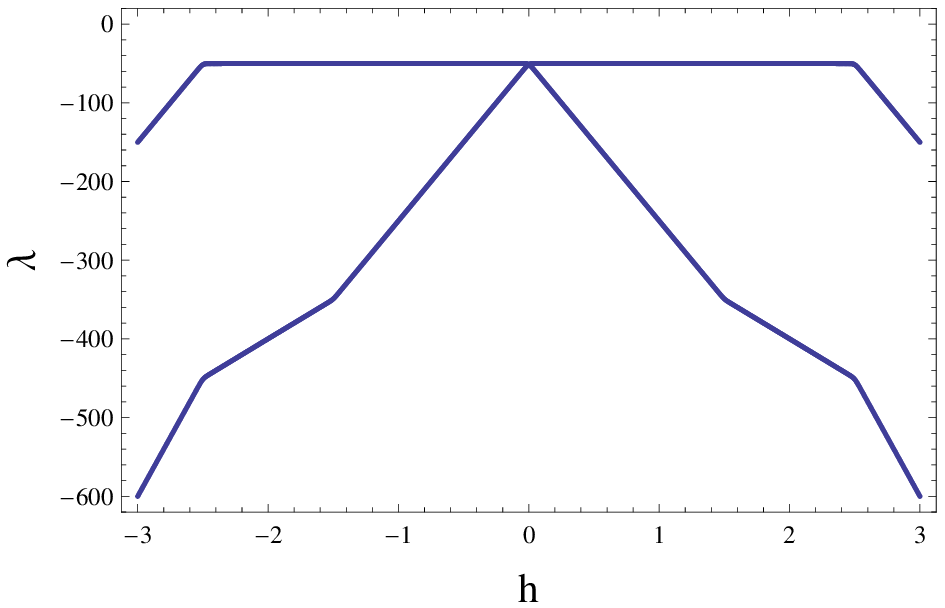} &
\includegraphics[width=8cm]{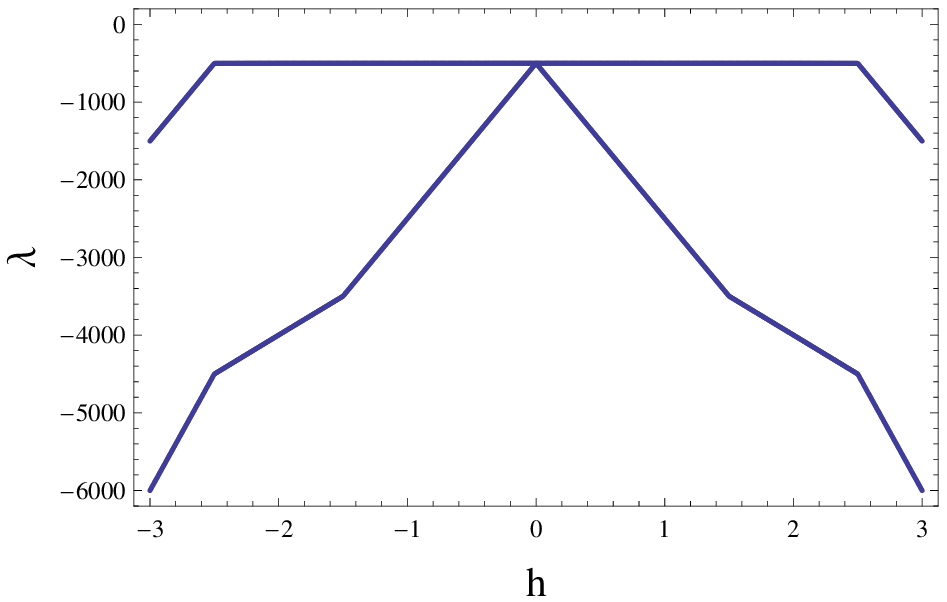}\\
\end{tabular}
\caption { \small{Plot of Lyapunov exponents at exchange parameters
$J=1.5$, $J_1=1$, $\Delta=1$ and temperature (a) $T=0.01$ (b)
$T=0.001$.}} \label{lyap2}
\end{center}
\end{figure}

\begin{figure}[!h]
\begin{center}
\begin{tabular}{cc}
{\small (a)}& {\small (b)}\\
\includegraphics[width=8cm]{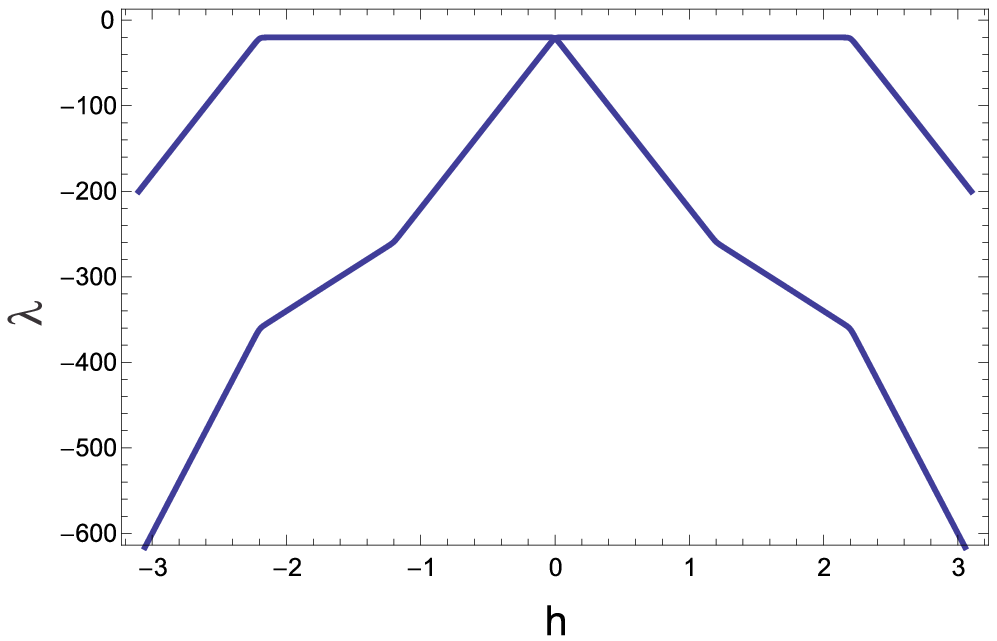} &
\includegraphics[width=8cm]{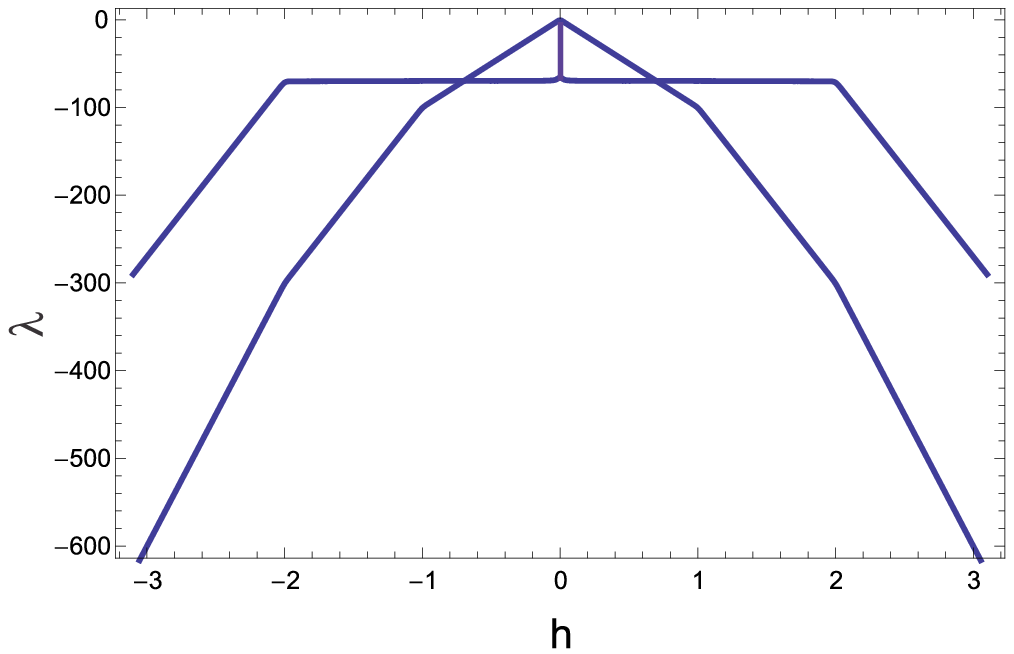}\\
\end{tabular}
\caption { \small{Plot of Lyapunov exponents for different
ferrimagnetic (FRI1 and FRI2) phases at exchange parameters $J=1.5$,
$J_1=1$ and the temperature $T=0.01$ (a) $\Delta=0.6$ (b)
$\Delta=-0.6$.}} \label{lyap3}
\end{center}
\end{figure}

where $J(x,y)$ is the Jacobian of the mapping (8). Expression
(\ref{lyapunov}) will let us count up the meanings of Lyapunov
exponents depending from an external magnetic field ($h$) at fixed
values of constants of interaction $(J, J_{1})$, the anisotropy
parameter ($\Delta$) and temperature ($T$). Figure \ref{lyap1} shows
the dependence of Lyapunov exponents on an external magnetic field
for the antiferromagnetic and ferromagnetic cases. As it can be seen
from Fig. \ref{lyap1} (a) both values of Lyapunov exponent are equal
to each other at absence of an external magnetic field. Another
interesting property of Lyapunov exponent is the existence of the
plateau on maximal Lyapunov exponent curve for the antiferromagnetic
case. It is important to mention that the locations of plateaus on
the maximal Lyapunov exponent curve (Fig. \ref{lyap1} (a)) coincide
with the locations of plateaus on the magnetization curve (Fig.
\ref{mag1} (b)). Such an interesting phenomena of Lyapunov exponent
has also been observed by investigating two, three and six spin
exchange interactions Heisenberg model on kagome lattice in an
external magnetic field \cite{Hrach}. Figure \ref{lyap1} (b) shows
that for the ferromagnetic case the maximum value of the maximal
Lyapunov exponent tends to zero.

In Fig.  \ref{lyap2} we show the behavior of Lyapunov exponent for
the antiferromagnetic case at lower temperature. As it can be seen from
figure the absolute values of Lyapunov exponents are increasing by
decreasing the temperature.

Next, we turn our attention to the behavior of Lyapunov exponent
curves in different ferrimagnetic (FRI1 and FRI2) phases. For this
purpose, Lyapunov exponent curves for fixed values of interaction
constants and the absolute temperature are plotted in Fig. (6). At
low temperatures, the minimum and maximum Lyapunov exponents have
different behavior. Only at $h =0$ the maximum value of the minimum
Lyapunov exponent equal to value of the maximum one, when the system
is in FRI2 phase (Fig. 6 (a)), and there is no intersection of
Lyapunov exponents for $h \neq 0$. There is a super stable point
($\lambda_{max} \rightarrow\infty$), when $h=0$ and at $T
\rightarrow \infty$ in FRI2 phase. There are two points of
intersections for the maximum and minimum values of Lyapunov
exponents and coincide when $h = 0$ in FRI1 phase (Fig. 6 (b)).
Lyapunov exponents are tending to zero in thermodynamic limit ($T
\rightarrow 0$) in FRI1 phase at absence of an external magnetic
field.

Now let us investigate another interesting property of the recursion
relation (8), namely superstability. First of all we will define
superstability for the one dimensional recursion relation. Generally
one dimensional recursion relation $x_{n}=f(x_{n-1})$ is said to be
superstable if the following relation takes place
\cite{Kaneko,Howard,Ott,Chakhmakhchyan}
\begin{eqnarray}
\frac{d^{n}f(x^{\ast})}{dx}=0,
\end{eqnarray}
where $x^{\ast}$ is the fixed point of $f(x)$. An other way
superstability can be defined by using definition of Lyapunov
exponent. The system is superstable when
\begin{eqnarray}
\lambda=\lim_{n\rightarrow\infty}{\frac{1}{n}}
\ln\mid{\frac{df^{n}(x^{\ast})}{dx}}\mid=-\infty.
\end{eqnarray}
In the same way we can define superstability for two dimensional
recursion relations (8). As it is already mentioned above for the
antiferromagnetic case the absolute values of Lyapunov exponents are
increasing by decreasing the temperature. Putting values of the
exchange parameters ($J=1.5$, $J_1=1$ and $\Delta=1$) into equation
(17) we can see that at thermodynamic limit at absence of an
external magnetic field the following relation takes place for
Lyapunov exponents
\begin{eqnarray}
\lim_{T\rightarrow0}\lambda_{1}=\lambda_{2}=-\infty,
\end{eqnarray}
which shows the existence of the super stable point.

We have analyzed the behavior of the magnetization for the
spin-$\frac{1}{2}$ Ising-Heisenberg model on diamond chain for
different values of anisotropy parameter $\Delta$. For the
antiferromagnetic case ($J>0$, $J_1>0$) at fixed values of exchange
parameters $J=1.5$ and $J_1=1$, the temperature T and for $\Delta
<\frac{1}{3}$ the magnetization curves have the same appearance as
in Fig. 2 (e). The values of characteristic Lyapunov exponent tend
to zero at absence of an external magnetic field and at $T
\rightarrow 0$, which means that there is no supercritical behavior
for the antiferromagnetic case when $\Delta <\frac{1}{3}$.

For the antiferromagnetic case ($J>0$, $J_1>0$) at low temperatures
and for values of the anisotropy parameter $\Delta$ ($\Delta
>\frac{1}{3}$) the magnetization curves have the same behavior as
shown in Fig. 2 (b). The changes of the anisotropy parameter
$\Delta$ only brings to the changes of the width of the
magnetization plateau at one third. The values of characteristic
Lyapunov exponent tend to $-\infty$ at absence of an external
magnetic field and at $T \rightarrow 0$, which means that there is a
superstable point for the antiferromagnetic case for positive values
of anisotropy parameter $\Delta$. Usually a super stable point lies
between bifurcation points \cite{Kaneko,Howard,Ott,Chakhmakhchyan}.
In our case for the spin-$\frac{1}{2}$ Ising-Heisenberg model on a
diamond chain there are no bifurcation points but the maximal
Lyapunov exponent tends to minus infinity. So we get the phase
transition in the super stable point at $h=0$ and $T \rightarrow 0$.
For the first time we get the phase transition point at the super
stable one.

\section{Conclusion \label{Conclusion}}

By using the recursion relation technique, we have studied magnetic
properties of the exactly solvable spin-$\frac{1}{2}$
Ising-Heisenberg model on diamond chain. Recursion relation
technique allowed us to construct the exact two-dimensional
recursion relation for the partition function. The behavior of the
total magnetization with respect to its saturation value has been
investigated. The existence of the magnetization plateau at one
third of saturation value of magnetization has been observed in the
antiferromagnetic case. The ground-state phase diagrams in $\Delta -
h$ plane show the existence of two ferrimagnetic (FRI1 and FRI2)
phases and one paramagnetic (PRM) phase in the antiferromagnetic
case and one ferrimagnetic (FRI) and a paramagnetic (PRM) phases in
the ferromagnetic case.

The properties of Lyapunov exponents were also discussed. The
existence of the plateau of the maximal Lyapunov exponent curve was
observed at low temperatures. It was detected the different behavior
for Lyapunov exponent curves in two ferrimagnetic phases. We have
shown that for the antiferromagnetic case in the thermodynamic limit
($T\rightarrow0$) both values of Lyapunov exponent tend to $-\infty$
at absence of an external magnetic field which is corresponding to
the superstable point.

\section{Acknowledgments}

This work has been supported by the French-Armenian Grant No. CNRS
IE-017 and Marie Curie IRSES SPIDER, and project
PIRSES-GA-2011-295302 (NA). The authors are grateful to R. Kenna and
H. Lazaryan for useful discussions.

\end{document}